\documentclass[prc,twocolumn,epsfig,nofootinbib,floatfix]{revtex4}
\usepackage{graphics}
\usepackage{epsfig}
\usepackage{amsfonts}
\usepackage{amsmath}
\usepackage{bm}% bold math

\usepackage{color}

\begin{document}
\title{Monopole giant resonance in $^{100-132}$Sn, $^{144}$Sm and $^{208}$Pb}
\author{J. Kvasil$^{1}$,  D. Bo\v z\'ik$^{1}$, A. Repko$^{1}$, P.-G. Reinhard $^{2}$,
V.O. Nesterenko  $^{3}$, and W. Kleinig  $^{3,4}$}
\affiliation{$^{1}$\email{kvasil@ipnp.troja.mff.cuni.cz},
\it  Institute of Particle and Nuclear Physics, Charles University, Prague, Czech Republic}
\affiliation{$^{2}$  \it Technische
Universit\"at Dresden, Institut f\"ur Analysis, D-01062, Dresden, Germany}
\affiliation{$^{3}$   \it
Laboratory of Theoretical Physics, Joint Institute for Nuclear
Research, Dubna, Moscow region, 141980, Russia}
\affiliation{$^{4}$
  \it Institut f\"ur Theoretische Physik II,
     Universit\"at Erlangen, D-91058, Erlangen, Germany}

\pacs{24.30Cz, 21.60.Jz, 27.60.+j}

\begin{abstract}
The isoscalar giant monopole resonance (GMR) in spherical nuclei
$^{100-132}$Sn, $^{144}$Sm, and  $^{208}$Pb is investigated within
the Skyrme random-phase-approximation (RPA) for a variety of Skyrme
forces and different pairing options. The calculated GMR strength
functions are directly compared to the available experimental
distributions. It is shown that, in accordance to results of other
groups, description of GMR in Sn  and heavier Sm/Pb nuclei
needs different values of the nuclear incompressibilty, $K \approx$ 200
or 230 MeV, respectively. Thus none from the used Skyrme forces is able to
describe GMR  in these nuclei simultaneously. The GMR peak energy in
open-shell $^{120}$Sn is found to depend on the isoscalar effective
mass, which might be partly used for a solution of the above problem.
Some important aspects of the problem (discrepancies of available
experimental data, proper treatment of the volume and surface
compression in finite nuclei, etc) are briefly discussed.
\end{abstract}

\maketitle

\section{Introduction}
During the last decades, the giant monopole resonance (GMR) was subject of intense
studies, see e.g. \cite{Co08, Av13, StSt} for recent reviews and discussions. The GMR
provides a valuable information on the nuclear incompressibility \cite{Bl80} since its
energy centroid $E_{\rm GMR}$ can be directly related to the compression modulus
$K_{\rm A}$ within  a collective model \cite{Bl80},
\begin{equation}
E_{\rm GMR}^A = \sqrt{\frac{\hbar^2 K_{\rm A}}{m \langle r^2 \rangle_0}},
\label{1}
\end{equation}
where $A$ is the nucleus mass number, $m$ is the nucleon mass, and
$\langle r^2 \rangle_0$ is the ground-state mean-square radius.
 For nuclear matter, the commonly accepted value for
incompressibility is $K$ = 230-240 MeV, confirmed by relativistic as
well as non-relativistic mean field models \cite{Nik08}.

Despite many efforts, the  description of GMR still suffers from
some persisting problems. For example, the mean field
models with $K$ =230-240 MeV reproduce the GMR
experimental data in heavy and medium nuclei, like
$^{208}$Pb  and $^{144}$Sm \cite{Yo99,It03,Yo04}, but
fail to describe more recent GMR experiments for lighter
nuclei, like Sn and Cd isotopes
\cite{Lu04,Li07,Li10,Li12,Pa12}, which request a lower
incompressibility (see e.g. the discussion in
\cite{Co08, Av13,Ca12,Ve12}). In other words, none of
the modern self-consistent models, relativistic or
non-relativistic, can simultaneously describe the GMR in
all mass regions. This problem has been already analyzed from
different sides. It was shown that GMR centroids can be
somewhat changed by
varying the symmetry energy  at constant $K$ \cite{Co04}.
Different Skyrme forces and pairing options (surface,
volume, and mixed) were inspected \cite{Av13,Ca12,Ve12}.
Hartree-Fock-Bogoliubov (HFB)
and HF-BCS methods were compared \cite{Kh09}. All these
attempts have partly conformed the description
but not solved the problem completely.

The modern GMR experimental data are
mainly delivered by two groups: Texas A$\&$M University (TAMU)
\cite{Yo99,Yo04,Lu04} and Research Center for Nuclear Physics
(RCNP) at Osaka University \cite{It03,Li07,Li10,Li12,Pa12}. Both groups
use ($\alpha, \: \alpha^{\:'}$) reaction and multipole
decomposition prescription to extract the E0 contribution from the
cross sections. However, these groups provide noticeably different
results and this should be also taken into account in the analysis of
the above problem \cite{Av13,Kv14}.

The present paper is also devoted to the problem of the
simultaneous description of the GMR in Sm-Pb and Sn nuclei. For
this aim, the E0 strength functions are calculated within Skyrme RPA
\cite{Re92} and directly compared to the available experimental
data. Note that in some previous studies \cite{Av13,Kh09}
the GMR analysis was limited to inspection of energy
centroids calculated through sum rules. Such analysis can
be ambiguous as it depends on the choice of the energy
interval where the sum rules are calculated. Besides that, gross
structure of E0 strength in the GMR and around (e.g. a high-energy
essential tail of the E0 strength in the RCNP data
\cite{It03,Li07,Li10,Li12,Pa12}) escapes considerations.
In our
opinion, a detailed exploration should include a direct
comparison of the calculated and experimental strength
distributions, which is just done in the present study.

Our calculations are performed using the family of SV Skyrme forces
\cite{Kl09} which cover a wide range of the nuclear matter parameters
and thus are convenient for a systematic investigation.  Both surface
and volume pairing options are applied.  As shown below, our analysis
confirms that incompressibility K$\approx 230$ MeV suits to describe
E0 strength in heavier nuclei ($^{208}$Pb and $^{144}$Sm) while
lighter Sn isotopes need lower values K$\approx 200$ MeV. In
connection to this problem, a significant discrepancy between TAMU and
RCNP experimental data is discussed. Besides we inspect dependence of
the GMR peak energy on various nuclear matter variables in three
representative Sn isotopes: neutron-deficit doubly magic $^{100}$Sn,
stable semi-magic $^{120}$Sn, and neutron-rich doubly magic $^{132}$Sn

The paper is organized as follows.  In Sec.~II, the theoretical
framework and calculation details are sketched. In Sec.~III, the
results are discussed. Further the summary is done.

\section{Theoretical framework}

The calculations of the E0 strength function are performed within
RPA based on the Skyrme energy functional (see e.g. \cite{Kl09})
\begin{equation}
\mathcal{E}(\rho,\tau, \vec{J}, \vec{j}, \vec{\sigma}, \vec{T}) =
\mathcal{E}_{\rm{kin}} + \mathcal{E}_{\rm{Sk}} +
\mathcal{E}_{\rm{Coul}} + \mathcal{E}_{\rm{pair}}  \label{6}
\end{equation}
depending on a couple of local densities  ($\rho(\vec{r})$ - nucleon,
$\tau(\vec{r})$ - kinetic energy, $\vec{J}(\vec{r})$ - spin-orbit,
$\vec{j}(\vec{r})$ - current, $\vec{\sigma}(\vec{r})$ - spin,
$\vec{T}(\vec{r})$ - vector kinetic energy). Here
$\mathcal{E}_{\rm{kin}}$, $\mathcal{E}_{\rm{Sk}}$,
$\mathcal{E}_{\rm{Coul}}$ and $\mathcal{E}_{\rm{pair}}$ are
kinetic energy, Skyrme, Coulomb, and pairing terms, respectively.
The explicit expressions for these terms are found elsewhere, see e.g.
\cite{Kl09}.

Using Hartree-Fock (HF) with BCS treatment of the pairing
we obtain the effective Hamiltonian as a sum of the
quasiparticle mean field $\hat{h}_{\rm{HF+BCS}}$ and
residual interaction $\hat{V}_{\rm{res}}$ \cite{Nes02,Kl09,ReKv}:
\begin{equation}
\hat{H} = \hat{h}_{\rm{HF+BCS}} + \hat{V}_{\rm{res}} \label{7}
\end{equation}
where
\begin{equation}
\hat{h}_{\rm{HF+BCS}} = \int d^3r \: \sum_{d+} \frac{\delta
\mathcal{E}}{\delta J_{d+}(\vec{r})} \: \hat{J}_{d+}(\vec{r}) \; ,
\label{8}
\end{equation}
\begin{equation}
\hat{V}_{\rm{res}} = \frac{1}{2} \sum_{d, d'} \int d^3r  \int
 d^3r' \frac{\delta^2 \mathcal{E}}{\delta J_d
 \delta \! J_{d^{'}}}  : \hat{J}_d(\vec{r})
\hat{J}_{d^{'}}(\vec{r}^{\:'}) : \; . \label{9}
\end{equation}
The $\hat{h}_{\rm{HF+BCS}}$ involves only time-even
densities ($\rho,\: \tau, \: \vec{J}$)
while in $\hat{V}_{\rm{res}}$ all the densities are embraced.
The symbol $:\;:$ in (\ref{9}) means the normal product of the
involved operators with respect to the quasiparticle creation
$\alpha_i^+$ and annihilation $\alpha_i$ operators \cite{ReKv}.
Then, using the standard RPA procedure, we obtain the RPA equation
%\cite{ReKv}
\begin{equation}
\left(
\begin{array}{cc}
A  &  B  \\
B^* &  A^*
\end{array}
\right)
\left(
\begin{array}{c}
c^{(\nu-)} \\
c^{(\nu+)}
\end{array}
\right) =
\left(
\begin{array}{cc}
E_{\nu}  &  0  \\
0 &  -E_{\nu}
\end{array}
\right)
\left(
\begin{array}{c}
c^{(\nu-)} \\
c^{(\nu+)}
\end{array}
\right)
\label{10}
\end{equation}
for the two-quasiparticle (2qp) forward and backward amplitudes
$c^{(\nu\pm)}$ of the phonon creation operator
\begin{equation}
Q_{\nu}^+(\lambda \mu) = \sum_{i\geq j} C_{j_i m_i j_j
m_j}^{\lambda \mu} \left( c^{(\nu-)}_{ij} \alpha_i^+ \alpha_j^+ -
c^{(\nu+)}_{ij} \alpha_{\bar{j}} \alpha_{\bar{i}} \right) . \label{11}
\end{equation}
Here $C_{j_i m_i j_j m_j}^{\lambda\mu}$ are Clebsch-Gordan
coefficients, $\nu$ numerates RPA states,  $E_{\nu}$ is the RPA
energy, $A$ and $B$ stand for the RPA matrices with the elements
\begin{eqnarray}
&&
A_{ijkl} = \delta_{ij,kl} \: \epsilon_{ij} +
\\
&& \sum_{dd^{'}} \frac{(-1)^{l_j + l_l}}{2 \lambda+1} \!
\int_{0}^{\infty} \!\!\!\! dr r^2 \: \frac{\delta^{\:2}
\mathcal{E}}{\delta J_d \: \delta \! J_{d^{'}}} \:
J_{d;ij}^{(\lambda)}(r) \: J_{d^{'};kl}^{\:*\:(\lambda)}(r) \; ,
\nonumber \\[0.5cm]
&&
B_{ijkl} =
\\
&& \sum_{dd^{'}} \frac{\gamma_d (-1)^{l_j + l_l}} {2 \lambda+1}
\! \int_{0}^{\infty} \!\!\!\! dr r^2 \: \frac{\delta^2
\mathcal{E}}{\delta J_d \: \delta \! J_{d^{'}}} \:
J_{d;ij}^{(\lambda)}(r) \: J_{d^{'};kl}^{\:*\:(\lambda)}(r) \nonumber
\label{12}
\end{eqnarray}
where $\epsilon_{ij}$ are 2qp energies, $\gamma_d$=+1 for
time-even and -1 for time-odd densities. Further,
$J_{d;ij}^{(L)}(r)$ are radial parts of decompositions
\cite{ReKv}
\begin{eqnarray}
&&
\hat{J}_d(\vec{r}) =
\sum_{ijLM}^{i>j} J^{(L)}_{d;ij}(r) \: \frac{(-1)^{l_i+L+1}}{\sqrt{2\lambda+1}}
\\
&& \qquad \qquad C_{j_i m_i j_j m_j}^{L M} \: Y_{L M}(\hat{r}) \:
\left( \alpha^+_{\bar{i}} \alpha^+_{\bar{j}} - \gamma_d \alpha_j
\alpha_i \right) \; , \nonumber
\end{eqnarray}
where $Y_{LM}(\hat{r})$ are spherical harmonics (more complicated
radial parts for the vector and tensor densities can be found in
\cite{ReKv}). Diagonalization of the RPA matrix
gives amplitudes $c_{ij}^{(\nu\pm)}$ and phonon energies
$E_{\nu}$.

By using the structure and energies of one-phonon
states, the strength function
\begin{equation}
S(E0,\:E) = \sum_{\nu} \left| \:\langle \nu \:|\:
\hat{M}(E0) \:|\: 0 \rangle \right|^2 \:
\xi_{\Delta}(E-E_{\nu}) \label{15}
\end{equation}
for the monopole transition operator $\hat{M}(E0)=r^2Y_{00}$
is calculated.  The Lorentz weight
\begin{equation}
\xi_{\Delta}(E-E_{\nu}) = \frac{1}{2 \pi}
\frac{\Delta}{(E-E_{\nu})^2 + \frac{\Delta^2}{4}}
\label{14}
\end{equation}
with the averaging parameter $\Delta$= 2 MeV is used. Such
averaging is found optimal for a simulation of the
smoothing effects beyond RPA (escape widths and coupling to
complex configurations). This then allows a convenient comparison
of the calculated and experimental strengths.

\begin{figure*}[t]
%\begin{minipage}{18pc}
\includegraphics[width=27pc]{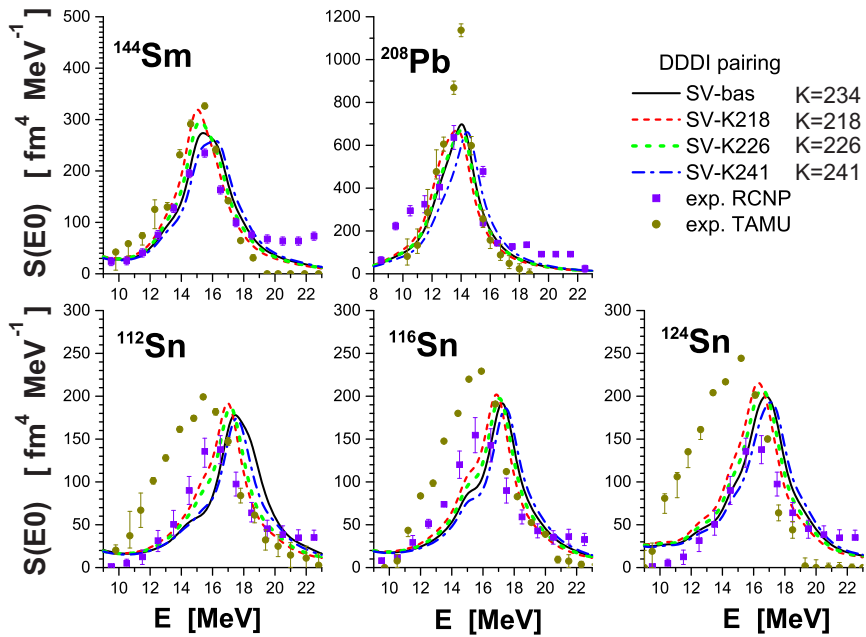}
\caption{\label{fig1}
E0(T=0) strength functions in $^{144}$Sm, $^{208}$Pb and
$^{112,116,124}$Sn, calculated with SV forces
SV-K218, SV-K226, SV-bas, and SV-K241. The incompressiblity moduli
K are indicated for each force in MeV. In Sm and Sn, the DDDI pairing
is used. The Lorentz averaging parameter is $\Delta$ = 2 MeV.
The results are compared with TAMU data for Sm/Pb \cite{Yo99,Yo04},
and Sn \cite{Lu04} and RCNP data for Sm \cite{It03}, Pb \cite{Uc04},
and Sn \cite{Li07}.}
\end{figure*}

The calculations exploit a set of Skyrme SV forces \cite{Kl09},
consisting of groups of parameterizations with varying one of
the  nuclear matter variables: incompressibility modulus $K$,
isoscalar effective mass $m^*_0/m$, TRK sum-rule enhancement
$\kappa$ (related to the isovector effective mass), and the
symmetry energy  $a_\mathrm{sym}$. Thus SV forces are convenient
for systematic investigation of the dependence of the results on
the basic nuclear matter features. For the comparison, some other
Skyrme parameterizations with essentially different incompressibilities
are used: SkP$^{\delta}$ (K = 202 MeV) \cite{SkPd}, SKM$^*$ (K = 217 MeV)
\cite{SkMs}, SLy6 (K = 230 MeV) \cite{SLy6}, and SkI3 (K = 258 MeV) \cite{SkI3}.

In our strength distributions, the spurious mode related to the pairing-induced
non-conservation of the particle number lies at 4-6 MeV, which is safely
below the GMR peaked at 14-16 MeV. To minimize the impact of the spurious
mode, only the strength at the excitation energy $E >$ 9 MeV is considered.

The calculations use a radial coordinate-space grid with step size 0.1 fm. The
large configuration space involving 2qp spectrum
up to 140 MeV is exploited. The energy weighted sum rule EWSR(E0)$=
\frac{\hbar^2}{2 \pi m}A \langle r^{2} \rangle_0$ estimated at the
energy interval 9-45 MeV is exhausted  by 98-105\%. The excess of EWSR
arises in nuclei with pairing due to the remaining weak tail
of the spurious mode.

Proton and neutron pairing are taken into account
in semi-magic $^{144}$Sm and $^{112,116,120,124}$Sn, respectively.
The pairing potential reads
\begin{equation}
V_{pair} (\vec{r}, \vec{r}^{\:'}) =
V_{0,q} \: \Big[ 1 - \eta \Big(\frac{\rho(\vec{r})}{\rho_0}\Big)
\: \Big] \: \delta(\vec{r} - \vec{r}^{\:'}) ,
\label{5}
\end{equation}
where $q$ stands for protons or neutrons.
The $V_{0,p}$ and $V_{0,n}$ are the pairing strengths, $\rho(\vec{r})$ is the nucleon density and $\rho_0$ is
the density of symmetric nuclear matter at equilibrium. The parameter $\eta$
switches the pairing options between the volume
delta-interaction (DI) for $\eta$=0 and surface density-dependent
delta-interaction (DDDI) for $\eta$=1. Both options are used
in the calculations. Pairing is treated at the BCS level. This means that,
at each HF iteration for single-particle
wave functions, the Bogoliubov coefficients $u_i$ and $v_i$
are determined within the BCS and then
introduced to the Skyrme densities. However here,
unlike the HFB case, the HF iterations exploit the single-particle
hamiltonian without the pairing term  \cite{Bender}.
The RPA calculations take into account
the pairing particle-particle channel.

\begin{figure*}[t]
%\begin{minipage}{18pc}
\includegraphics[width=27pc]{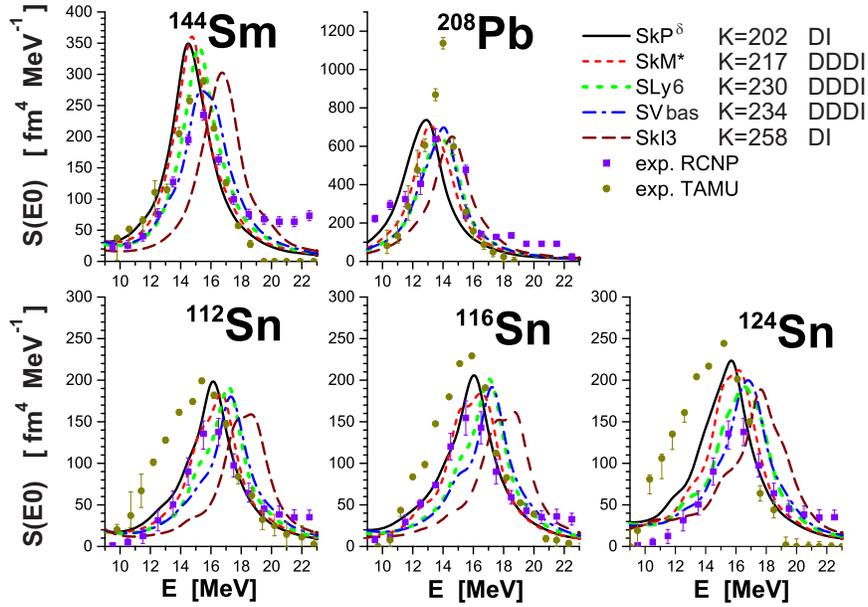}
\caption{\label{fig2}
The same as in Fig. 1 but for the Skyrme forces SkP$^{\delta}$,
SkM$^{*}$, SLy6, SV-bas, and SkI3. The incompressiblity moduli
K (in MeV) and pairing options (DI or DDDI) are indicated for each
force.}
\end{figure*}

\section{Results and discussion}

Figure 1 shows the RPA results for isoscalar (T=0) GMR in
doubly-magic $^{208}$Pb, neutron semi-magic $^{144}$Sm,
and proton semi-magic $^{112,116,124}$Sn.
The SV parameterizations with different values of the
incompressibility $K$  (as indicated at the figure)
are used. The calculated strength functions are compared
with TAMU \cite{Yo99,Yo04,Lu04} and RCNP \cite{It03,Uc04,Li07}
experimental data. For the convenience of comparison, the
TAMU data, being initially presented in units of the fraction
of the EWSR, are transformed to units fm$^4 \: \rm{MeV}^{-1}$
used by RCNP. In the calculated strength function, the Lorentz
averaging parameter $\Delta$=2 MeV is used as most convenient
for the comparison between the calculated and RCNP strengths.
Such averaging produces GMR amplitudes and widths similar
to the RCNP ones. Then the actual model output is reduced to the
GMR shape, integral strength, and energy.
As seen from Fig. 1, the calculated GMR integral strength is
close to the RCNP one expressed in the same units.

Before further comparison of the theory and experiment,
the significant discrepancy between RCNP
and TAMU data should be discussed.  Fig. 1 shows a large
difference at the left wing of the GMR strength. In principle, this difference
can be
reduced by a proper re-scaling of the TAMU data.
However, scaling cannot conceal two other apparent differences.
First, as compared to RCNP, TAMU gives somewhat lower
energy position of the GMR peak in Sm and Sn. Second, the
RCNP data exhibit a long uniform
tail above the GMR. Only onset of this tail at 19 MeV $< E
<$ 23 MeV is seen in Fig. 1, though actually it continues at
least up to 33 MeV \cite{Li07}. Such a tail is absent in
TAMU data, see discussion \cite{Yo04}. This
difference can affect determination of the experimental GMR
centroids which are usually estimated through the sum
rules and thus depend on the chosen energy intervals.
Altogether, these two RCNP/TAMU discrepancies should
be taken into account in the comparison of the calculated GMR
with the experiment.
\begin{figure*}[t]
%\begin{minipage}{18pc}
\includegraphics[width=25pc]{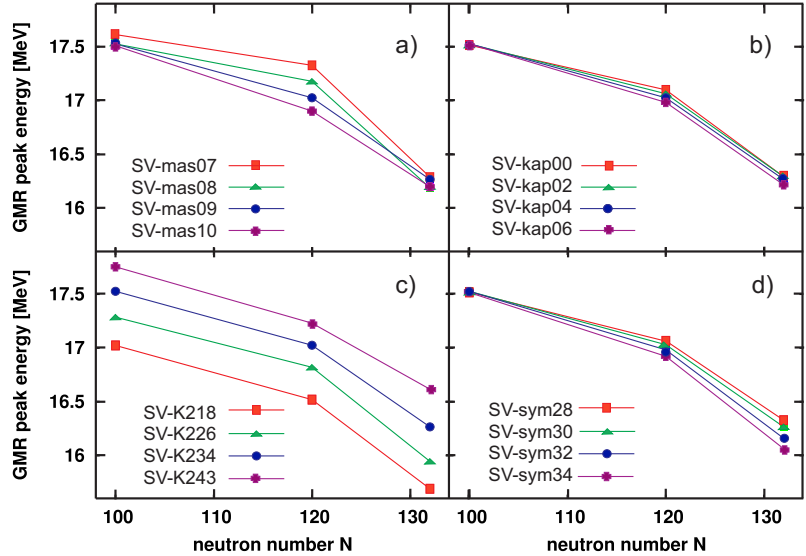}
\caption{\label{fig3} Dependence of the GMR peak energies in
$^{100,120,130}$Sn on the nuclear matter values: (a)  isoscalar
effective mass ($m^*_0$/m=0.7, 0.8, 0.9, 1.0), b) isovector
enhancement factor ($\kappa$=0, 0.2, 0.4, 0.6), c)
incompressibility modulus ($K$ =218, 226, 234, 243 MeV)
and d) symmetry energy($a_\mathrm{sym}=$28, 30, 32, 34 MeV). The
RPA calculations \protect\cite{Re92} are performed with the family
of SV Skyrme forces \protect\cite{Kl09}.}
\end{figure*}

Figure 1 also shows that, in accordance to previous studies (see
\cite{Co08, Av13} and references therein), the calculate energy
position of the GMR noticeably depends on the the incompressibility $K$.
The larger $K$, the higher GMR. For instance, in $^{144}$Sm the
force SV-K241 with $K$ = 241 MeV gives the GMR maximum at 16.2 MeV
while SV-K218 with $K$ = 218 MeV downshifts this maximum to 15 MeV.
It is also seen that the forces with a large K, like SVbas,
provide  a good agreement with TAMU and RCNP data in heavy
nuclei $^{144}$Sm and $^{208}$Pb but certainly overestimate the GMR energy
in lighter Sn isotopes. The description of GMR in Sn
needs forces with a much smaller incompressibility (even the force SV-K218
is not enough). In other words, Sn isotopes demonstrate a remarkable softness
to the compression and this is valid for both RCNP and TAMU data.

These conclusions are corroborated in Fig. 2 where Skyrme
parameterizations beyond the SV-family (SkP$^{\delta}$ \cite{SkPd},
SkM$^{*}$\cite{SkMs}, SLy6 \cite{SLy6}, SkI3 \cite{SkI3})
with a broader incompressibility range $K$=202 $\div$ 258 MeV are applied.
Fig. 2  shows that the best description of the
GMR energy $E_{\rm{GMR}}$ in Sn isotopes is obtained with the low-$K$
SKP$^{\delta}$ and SkM* forces
but these forces noticeably underestimate the GMR energy in
$^{144}$Sm and $^{208}$Pb. For the latter nuclei, the
parametrization SV-bas with a large $K$=234 MeV is best but it
fails in Sn isotopes. The force SkI3 with the highest incompressibility
$K$ = 258 MeV overestimates $E_{\rm{GMR}}$ in all the considered nuclei. Both
pairing options, DI and DDDI, give similar results.

It is instructive to compare sensitivity of the centroid energy $E_{\rm{GMR}}$ to $K$ and
also to other nuclear matter variables.
Fig. 3 shows the dependence of $E_{\rm{GMR}}$ on the isoscalar effective mass $m^*_0/m$,
isovector enhancement factor $\kappa$, incompressibility $K$, and symmetry energy  $a_\mathrm{sym}$ for
neutron-deficit doubly-magic $^{100}$Sn, stable semi-magic $^{120}$Sn, and neutron-rich doubly-magic
$^{132}$Sn. One finds that $E_{\rm{GMR}}$ indeed depends most strongly on $K$: the difference in
$E_{\rm{GMR}}$ for $K$=218  and 243 MeV reaches 1 MeV. The $E_{\rm{GMR}}$ decreases from $^{100}$Sn to
$^{132}$Sn in general agreement with the empirical estimation $E^{\rm e}_{\rm{GMR}}\approx 78 A^{-1/3}$ MeV
\cite{Harbook}. The dependence of $E_{\rm{GMR}}$ on $\kappa$ and $a_\mathrm{sym}$ is
generally weak (for exception of the neutron-rich nucleus $^{132}$Sn). However $^{120}$Sn
demonstrates a significant dependence on the isoscalar effective mass: the less $m^*_0/m$, the
higher $E_{\rm{GMR}}$. This is because decrease of $m^*_0/m$ makes the single-particle spectrum more dilute
and thus results in higher $E_{\rm{GMR}}$. In $^{120}$Sn, this effect seems to be enhanced by the neutron
pairing. Dependence on $m^*_0/m$ can be used for a further conformance of the GMR description in Pb/Sm and
Sn nuclei.

Finally note that the problem of the simultaneous description of the GMR in Pb/Sm and Sn
can have various reasons. Perhaps available Skyrme parameterizations
and paring treatments are not yet optimal enough. It is also possible
that accuracy of the Skyrme-like energy density functionals
have already reached their limits (see discussion \cite{Kor14}) and
some essential modifications of the functionals,
e.g. a richer density dependence \cite{Sto10},
are in order. Note that there are also other cases when
Skyrme forces cannot simultaneously describe  nuclear
excitations in closed- and open-shell nuclei. For example, none Skyrme
parametrization can simultaneously reproduce the spin-flip M1 GR
in closed-shell spherical $^{208}$Pb and open-shell rare-earth
deformed nuclei \cite{Ve09,Ne10}.
More versatile density dependence would also allow to distinguish
between the bulk incompressibility $K$ and surface incompressibility
 $K_A^{\rm{sur}}$.
Note that the incompressibility in finite
nuclei can be different at the nuclear surface
and interior because these regions have different nucleon densities. Besides,
at the nuclear surface, the impact of pairing is most essential. So the separate
consistent estimations of the volume $K_A^{\rm{vol}}$ (to be associated with the nuclear matter
incompressibility $K$) and surface $K_A^{\rm{sur}}$
incompressibility might be useful, see analysis of the leptodermous expansion
of $K_A$ in \cite{StSt}.

As mentioned above, the discrepancy between RCNP and TAMU experimental
data also hampers solution of the problem.  Note that RCNP/TAMU data
deviate also in deformed nuclei. For example, TAMU gives (in agreement
with recent Skyrme RPA calculations \cite{Kv14}) a clear two-bump GMR
structure in $^{154}$Sm \cite{Yo04}, explained by the
deformation-induced coupling between E0(T=0) and E2(T=0) modes.
Instead RNCP data give in this nucleus a one-bump GMR structure
\cite{It03}.  In this connection, note that TAMU and RCNP experiments
use $\alpha$-particle beams with different incident energies. Since
($\alpha', \alpha$) is a peripheral reaction, the TAMU and RCNP
experiments can probe different surface slices and thus exhibit
different compression responses which may be resolved by
distinguishing bulk and surface incompressibility $K_A^{\rm{vol}}$ and
$K_A^{\rm{sur}}$.  Certainly these discrepancies call for more
accurate measurements and analysis of the GMR.

\section{Conclusions}

The giant monopole resonance (GMR) was explored in  $^{100-132}$Sn, $^{144}$Sm,
and  $^{208}$Pb in the framework  of the Skyrme random-phase-approximation (RPA)
with different Skyrme forces and pairing options. The calculations confirmed
results of numerous previous studies \cite{Co08} that GMR in $^{208}$Pb ($^{144}$Sm)
and  Sn isotopes
cannot be simultaneously described with one and the same  Skyrme parametrization.
The analysis calls for more accurate experiments and
matching the TAMU and RCNP experimental data.

Dependence of
GMR peak energies on the nuclear matter variables (incompressibility $K$,
isoscalar effective mass $m^*_0$/m, isovector enhancement factor $\kappa$,  and
symmetry energy  $a_\mathrm{sym}$) was examined in $^{100,120,132}$Sn.  The calculations
confirmed the well-known strong dependence of the results on $K$. Besides, a
sensitivity to the isoscalar effective mass $m^*_0$/m was revealed.
The latter opens an additional  window for further arrangement of the problem of the
simultaneous GMR description in Sm/Pb and Sn nuclei. Some
possible next steps in solution of this problem (further experimental progress,
proper treatment of the ratio between the volume
and surface incompressibility in finite nuclei, etc) were briefly discussed.

\section*{Acknowledgments}
The work was partly supported by the DFG grant RE 322/14-1, Heisenberg-Landau
(Germany-BLTP JINR), and Votruba-Blokhintsev (Czech Republic-BLTP JINR)
grants. P.-G.R. and W.K. are grateful
for the BMBF support under the contracts 05P12RFFTG and
05P12ODDUE, respectively. The support
of the Czech Science Foundation (P203-13-07117S) is appreciated.


\begin{thebibliography}{99}
\bibitem{Co08} %1
  Colo G 2008 {\it Phys. Part. Nucl.} {\bf 39} 286
\bibitem{Av13} %2
  Avogadro P and Bertulani C A 2013 {\it Phys. Rev.} C {\bf 88} 044319
\bibitem{StSt} %3
  Stone J R, Stone N J and Moszkowski S A 2014 arXiv: 1404.0744 [nucl-th]
\bibitem{Bl80} %4
  Blaizot J 1980 {\it Phys. Rep.} {\bf 64} 171
\bibitem{Nik08} %5
  Niksi\'c T, Vretenar D, and Ring P 2008 {\it Phys. Rev.} C {\bf 78} 034318
%--------exper Sm -Pb
\bibitem{Yo99} % 6 TAMU Sm, Pb
  Youngblood D H, Clark H L and Lui Y-W 1999 {\it Phys. Rev. Letters} {\bf 82}(4) 691
\bibitem{It03} % 7 RCNP Sm
  Itoh M et al 2003 {\it Phys. Rev.} C {\bf 68} 064602
\bibitem{Uc04} %  8 RCNP Pb
 Uchida M, et al. 2004 {\it Phys. Rev.} C {\bf 69} 051301(R)
\bibitem{Yo04} % 9 TAMU Sm, Pb
  Youngblood D H et al 2004 {\it Phys. Rev.} C {\bf 69} 034315
%----exp Sn, Cd
\bibitem{Lu04} % 10 TAMU Sn
  Lui Y-W, Youngblood D H, Tokimoto Y, Clark H L and John B
  2004 {\it Phys. Rev.} C {\bf 70} 014307
\bibitem{Li07} % 11 RCNP Sn
  Li T et al 2007 {\it Phys. Rev. Letters} {\bf 99} 162503
\bibitem{Li10} % 12 RCNP Sn
  Li T et al 2010 {\it Phys. Rev.} C {\bf 81} 034309
\bibitem{Li12}  % 13 RCNP Sn
  Li T et al 2012 {\it Phys. Rev.} C {\bf 99} 162503
\bibitem{Pa12} % 14 RCNP Sn, Cd
  Patel D et al 2012 {\it Phys. Letters} B{\bf 718} 447
%-------------pairing etc
\bibitem{Ca12} %15
  Cao L, Sagawa H and Colo G 2012 {\it Phys. Rev.} C {\bf 86} 054313
\bibitem{Ve12} %16
  Vesely P, Toivanen J, Carlsson B C, Dobaczewski J, Michel M and Pastore A
  2012 {\it Phys. Rev.} C {\bf 86} 024303
\bibitem{Co04} %17
Col\'o G, Giai N V, Meyer J, Bennaceur K and Bonche P,
2004 {\it Phys. Rev.} C {\bf 70}, 024307
\bibitem{Kh09} %18
  Khan E 2009 {\it Phys. Rev.} C {\bf 80} 011307(R); ibid {\bf 80} 057302
%----------------------------
\bibitem{Kv14} %19
Kvasil J, Nesterenko V O, Repko A,  Bozik D, Kleinig W and Reinhard P-G 2014
arXiv:1407.3108[nucl-th].
\bibitem{Kl09}
  Klupfel P, Reinhard P-G, Burvenich T J and Maruhn A
2009 {\it Phys. Rev.} C {\bf 79} 034310
\bibitem{Nes02}
%title={Separable {RPA} for self-consistent nuclear models},
Nesterenko V O, Kvasil J and Reinhard P.--G. 2002
{\it Phys. Rev. C } {\bf 66} {44307}.
\bibitem{ReKv}
  Repko A, Kvasil J, Nesterenko V O and Reinhard P-G
  (under preparation for publication)
\bibitem{Re92}
 Reinhard P-G 1992 {\it Ann. Physik} {\bf 1} 632
%----Skyrme forces
\bibitem{SkPd} %14
 Reinhard P-G et al 1999 {\it Phys. Rev.} C {\bf  60} 014316
\bibitem{SkMs} %17
  J. Bartel, P. Quentin, M. Brack, C. Guet, and H.-B. Haakansson
  1982 {\it Nucl. Phys.} A {\bf 386} 79
\bibitem{SLy6} %14
  E. Chabanat et al 1997
  %P. Bonche, P. Haensel, J. Meyer, and R. Schaeffer,
  {\it Nucl. Phys.} A {\bf  627} 710
\bibitem{SkI3} %18
  P.-G. Reinhard and F. Flocard 1995
  {\it Nucl. Phys.} A {\bf 584} 467
%----------------------------
\bibitem{Bender}
Bender M, Rutz K, Reinhard P-G and Maruhn J A 2000 {\it Eur.
Phys. J.} A {\bf 8} 59
\bibitem{Harbook}
  Harakeh M N and van der Woude A 2001 Giant Resonances
(Oxford: Clarendon)
\bibitem{Kor14}
 Kortelinen M et al 2014 arXiv:1410.8303[nucl-th]
\bibitem{Sto10}
 Stoitsov M et al 2010 {\it Phys. Rev.} C {\bf  82} 054307
\bibitem{Ve09}
  Vesely P, Kvasil J, Nesterenko V O, Kleinig W, Reinhard P-G, and Ponomarev V Yu
  2009 {\it Phys. Rev.} C {\bf 80} 031302(R)
\bibitem{Ne10}
 Nesterenko V O, Kvasil J, Vesely P, Kleinig W, Reinhard P-G
 and Pomomarev V Yu 2010
 {\it J. Phys.} G {\bf 37} 064034
\end{thebibliography}
\end{document}